\documentclass[twoside,12pt]{article}
\usepackage{graphicx}

\begin{document}

\centerline{\Large \bf Are the Tails of Percolation Thresholds Gaussians ?}

\bigskip

P.M.C. de Oliveira$^{1,2}$, R.A. N\'obrega$^2$ and D. Stauffer$^{1,3}$.

\bigskip

\noindent 1) Laboratoire de Physique et M\'ecanique des Milieux
H\'et\'erog\`enes\par

\'Ecole Sup\'erieure de Physique et de Chimie Industrielles\par

10, rue Vauquelin, 75231 Paris Cedex 05, France

\noindent 2) Instituto de F\'\i sica\par

Universidade Federal Fluminense\par

av. Litor\^anea s/n, Boa Viagem, Niter\'oi, Brasil 24210-340

\noindent 3) Institute for Theoretical Physics\par

Cologne University\par

D-50923 K\"oln, Euroland

\medskip

\noindent e-mails: pmco@if.uff.br; rafaella@if.uff.br; 
stauffer@thp.uni-koeln.de.

\medskip


\bigskip

Abstract: The probability distribution of percolation thresholds in finite
lattices were first believed to follow a normal Gaussian behaviour. With
increasing computer power and more efficient simulational techniques, this
belief turned to a stretched exponential behaviour, instead. Here, based on a
further improvement of Monte Carlo data, we show evidences that this question
is not yet answered at all.

\bigskip\bigskip

\begin{figure}[hbt]

\begin{center}
\includegraphics[angle=-90,scale=0.4]{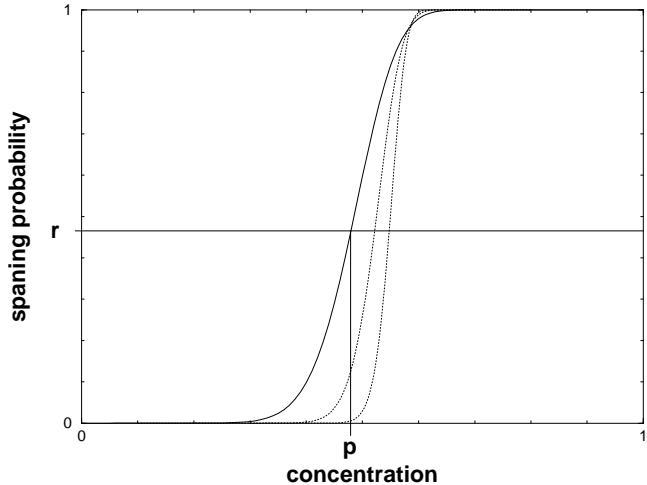}
\end{center}

\caption{Spanning probability function for a fixed lattice size $L$ (solid
line). For larger and larger sizes (dotted lines), this function approaches a
step. By fixing some value $r$ at the vertical axis, one can find a sequence of
values $p_L(r)$ at the horizontal axis approaching the critical threshold
$p_c$, for increasing lattice sizes.}

\end{figure}

	In reference \cite{NZ}, the percolation on a $N$-site square lattice is
treated with high numerical accuracy. Indeed, the best known estimate for the
critical threshold, $p_c = 0.59274621(13)$, comes from this work. In order to
study this kind of problems, the authors follow a very fruitful Monte Carlo
approach which allows one to obtain continuous functions of $p$, the
concentration of occupied sites, namely the canonical-like average

\begin{equation}
R(p)\,\, =\,\, \sum_n\,\, C_N^{\, n}\,\,\, p^n\, (1-p)^{N-n}\,\, R_n\,\,\,\, ,
\end{equation}

\noindent of some quantity $R$. Here, $R_n$ is a uniform average over all
configurations with just $n$ occupied sites, i.e. a microcanonical-like
average. $C_N^{\, n}$ are simple binomial factors. By filling the initially
empty lattice, site by site at random, and repeating this process many times,
one is able to get the discrete set of microcanonical averages $R_n$
accumulated into an $n$-histogram, over the entire range, $n$ = 0, 1, 2 $\dots$
$N$. From this set of numbers, the determination of the continuous $p$-function
$R(p)$ is straightforward.

	In particular the authors of \cite{NZ} fix attention at the horizontal
wrapping probability $R_L(p)$ around a $L \times L$ torus, i.e. a square
lattice with periodic boundary conditions on both directions. In the
thermodynamic limit, this function approaches a step: $R_\infty(p) = 0$ below
the critical threshold $p_c$, and $R_\infty(p) = 1$ above $p_c$. For finite
sizes, $R_L(p)$ presents a sigmoid aspect similar to figure 1. A good approach
to $p_c$ is to choose some fixed value $r$, and solve the equation $R_L(p) =
r$, getting the root $p$ shown at the horizontal axis. Here, one can appreciate
the advantage of knowing $R_L(p)$ as a {\bf continuous function} of $p$.
Keeping the same value $r$ and repeating this task for a series of increasing
lattice sizes (dotted lines), one gets a series of roots $p_{L_1}(r)$,
$p_{L_2}(r)$, $p_{L_3}(r)$ $\dots$ which converges to the desired threshold
$p_c$ in the thermodynamic limit. 

	The above reasoning is valid no matter which is the fixed value for $r$
one chooses. However, for the very particular choice $r^* = 0.521058290$, a
universal probability exactly known through conformal invariance arguments
\cite{Pinson}, the convergence becomes fast, i.e. the root $p(L)$ differs from
$p_c$ as $L^{-2-1/\nu} = L^{-2.75}$, where $\nu = 4/3$ is the correlation
length critical exponent. The above quoted accurate value for $p_c$ was
obtained in this way. For details, see \cite{NZ} and references therein.

	Reference \cite{bjp} proposes the mathematical form

\begin{equation}
p_L(r) = p_c + {1\over L^{1/\nu}} \Big[ A_0(r) + {A_1(r)\over L} +
{A_2(r)\over L^2} + \dots \Big]\,\,\,\, ,
\end{equation}

\noindent for estimators $p_L$ obtained from quantities like $R(p)$. The option
for the wrapping probabilities around the torus and the convenient choice of
Pinson's number $r = r^*$ lead to vanishing values for the two first terms
$A_0(r^*) = A_1(r^*) = 0$, a lucky coincidence which accelerates very much the
convergence.

	The quantity $R_L(p)$ is obtained, as quoted before, by filling up the
initially empty lattice site by site, at random. Clusters of neighbouring
occupied sites grow. As soon as the horizontal wrapping along the torus is set,
one books the corresponding value of $n$, the number of occupied sites so far,
and stops the process. For that particular sample, the wrapping probability
$R_L$ is a step function, i.e. $R_L = 0$ below $n$ and $R_L = 1$ above. The
same routine is repeated many times, in order to have a probability
distribution for $n$.  The various step functions are then superimposed to get
the microcanonical averages $R_n$ in equation (1), stored in an $n$-histogram.
Finally, the continuous canonical average $R_L(p)$ can be calculated for any
value of $p$.

	Each process of filling-up the lattice (one sample) yields a single
value $n$ for the statistics, i.e. just one more entry on the $n$-histogram. In
\cite{bjp}, we decided to improve this point, by changing the definition from
wrapping to {\bf spanning} probability, figure 1. We fix two parallel
horizontal lines on the $L \times L$ torus, separated by a distance of $L/2$,
for instance lines $i = 1$ and $i = 1+L/2$. The measured quantity is now the
probability of having these two lines connected by the same cluster of
neighbouring occupied sites, instead of the wrapping probability along the
whole torus. The advantage is that we can measure the same thing for lines $i =
2$ and $i = 2+L/2$, also for lines $i = 3$ and $i = 3+L/2$, and so on.
Moreover, also vertical parallel lines can be included into this counting. At
the end, from a single sample we store $L$ new entries into our $n$-histogram,
instead of just one more entry. Note that this advantage even increases for
larger and larger lattices.

	Within the same computational effort, our approach allows the test of
larger lattices. Because of that, we were able to confirm the validity of
equation (2) with high precision, by sampling 27 different lattice sizes from
$L = 18$ up to $L = 1594$, a 8-thousand factor in number of sites. On the other
hand, our definition does not allow the chance of vanishing both $A_0(r)$ and
$A_1(r)$ at once. We can have only $A_0(\hat{r}) = 0$, for a particular
universal probability $\hat{r} = 0.984786(11)$ numerically determined within
the same work \cite{bjp}. Independently, Cardy \cite{Cardy} tried to determine
it by conformal invariance arguments, however, in looking for configurations
which link two parallel lines, he was forced to disregard configurations which
also wrap along the other direction. As a result of using larger lattices but a
slower convergence rate of $L^{-1-1/\nu} = L^{-1.75}$, we get the same figure
$p_c = 0.59274621(33)$ \cite{bjp} as in \cite{NZ}, but within a 3 times larger
error bar.

\begin{figure}[hbt]

\begin{center}
\includegraphics[angle=-90,scale=0.4]{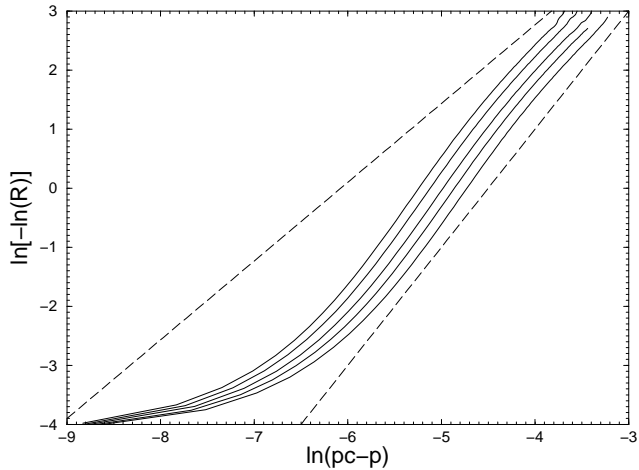}
\end{center}

\caption{Test of equations (3) or (4), for tails on the left of figure 1.  The
five continuous curves correspond to $L = 1594$, 1354, 1126, 958 and 802, from
left to right. In each case we sampled 4 million lattice-filling processes,
which corresponds to 6, 5, 5, 4, and 3 $\times 10^9$ entries in each
$n$-histogram, respectively. The statistics is improved by a factor over than
1000, compared with \cite{NZ} for equivalent lattice sizes. The dashed lines
show the alternative slopes 2 (right) or 4/3 (left). In the authors' opinion,
no definitive conclusion is possible.}

\end{figure}

\begin{figure}[hbt]

\begin{center}
\includegraphics[angle=0,scale=0.48]{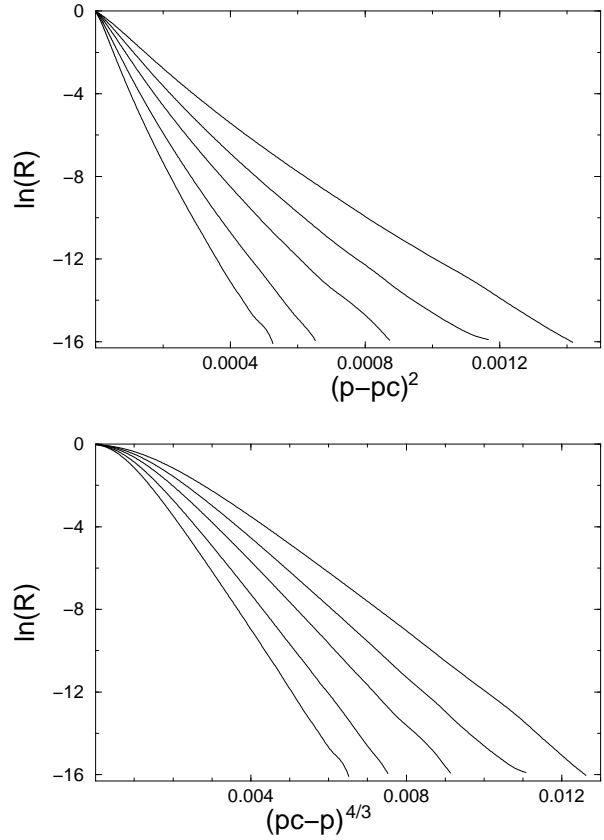}
\end{center}

\caption{Alternative test of equations (3) or (4), for the same lattice sizes
$L = 1594$, 1354, 1126, 958 and 802, from left to right.}

\end{figure}

	The non-Gaussian behaviour of the finite-lattice-threshold distribution
near the infinite-lattice critical point is already stablished \cite{Stauffer}. 
Here, we profit from the same simulational data in order to investigate the
distribution tails, {\bf far} from the critical point. Which is the
mathematical form of the tail observed in figure 1, below the root $p\,$? One
possible answer is a simple Gaussian form \cite{Levi,Wester}

\begin{equation}
R_L(p) \sim \exp[-K(p-p_c)^2]\,\,\,\, .
\end{equation}

\noindent Another alternative is a stretched exponential
\cite{Berlyand,Haas,Hovi}

\begin{equation}
R_L(p) = \exp[-C(p_c-p)^\nu]\,\,\,\, ,
\end{equation}

\noindent where the strict equality (for large $L$ and far from $p_c$) is a
consequence of the periodic boundary condition \cite{NZ}, which holds for our
data. Profiting from this strict equality, one can test equation (4) by
constructing a plot of $\ln[-\ln(R)]$ against $\ln(p_c-p)$. This was done in
\cite{NZ}, and we repeat the same for our data, in figure 2. Note that our
range for $\ln(p_c-p)$ (up to $-3.4$ for $L \sim 10^3$) is larger than in
reference \cite{NZ} (up to $-4.6$ for the same size). This means that we are
testing more deeply the distribution tails, thanks to our trick of sampling $L$
new entries for each run. Even so, the conclusion in favour of either equation
(3) or (4) is by no means obvious. Note a further difficulty in what concerns
equation (3), because the leading multiplicative constant in front of the
exponential is not necessarily 1.

	Another, perhaps better way to address the same question is by plotting
$\ln(R)$ twice, against $(p-p_c)^2$ and $(p_c-p)^{4/3}$. Figure 3 shows the
result for our data. Note that our range for $\ln(R)$ (down to $-16$) doubles
the one presented in \cite{NZ}. The would-be Gaussian case (up) presents clear
positive curvatures, whereas the would-be stretched exponential (down) presents
negative curvatures although not so pronounced. The exponent $4/3$ seems to fit
better, but one cannot extract a clear conclusion from these data.

	Still more undefined is the situation of the other tails on the right
of figure 1, above $p_c$. In this case (not shown), our accuracy limit for
$\ln(1-R)$ (down to $-16$) is reached much closer to $p_c$ than the case 
shown in figures 2 and 3, below $p_c$.

	Concluding, we present new Monte Carlo data concerning the distribution
probability of percolation thresholds on a finite square lattice. We address
the question of the mathematical form of the distribution tails, equation (3)
against (4). Even considering that our statistics is over than 1000 times
larger than previous works, no definitive conclusion can be extracted from our
data, in what concerns the asymptotic tail exponent.

\bigskip

Aknowledgements: This work is partially supported by Brazilian agencies FAPERJ 
and CNPq.


\begin{thebibliography} {99}

\bibitem{NZ} M.E.J. Newman and R.M. Ziff, {\it Phys. Rev.} {\bf E64}, 016706
(2001); {\it Phys. Rev. Lett.} {\bf 85}, 4104 (2000).

\bibitem{Pinson} H.T. Pinson, {\it J. Stat. Phys.} {\bf 75}, 1167 (1994).

\bibitem{bjp} P.M.C. de Oliveira, R.A. N\'obrega and D. Stauffer, {\it 
Braz. J. Phys.} {\bf 33}, 616 (2003), also in www.lanl.gov COND-MAT/0308525.

\bibitem{Cardy} J. Cardy, {\it J. Phys.} {\bf A35}, L565 (2002).

\bibitem{Stauffer} This point was not yet realized in D. Stauffer and A. 
Aharony, {\sl Introduction to Percolation Theory}, 2$^{\rm nd}$ edition, Taylor
and Francis, London (1992), but was corrected in the 1994 printing.

\bibitem{Levi} M.E. Levinshtein, B.I. Shklovskii, M.S. Shur and A.L. Efros, 
{\it Zh. \'Eksp. Teor. Fiz.} {\bf 69}, 386 (1975) [{\it Sov. Phys. JEPT}, {\bf 
42}, 197 (1976)].

\bibitem{Wester} F. Wester, {\it Int. J. Mod. Phys.} {\bf C11}, 843 (2000).

\bibitem{Berlyand} L. Berlyand and J. Wehr, {\it J. Phys.} {\bf A28}, 7127 
(1995).

\bibitem{Haas} U. Haas, {\it Physica} {\bf A215}, 247 (1995).

\bibitem{Hovi} J.-P. Hovi and A. Aharony, {\it Phys. Rev.} {\bf E53}, 235
(1996).

\end{thebibliography}
\end{document}